\documentclass{aastex}

\usepackage{emulateapj5}

\usepackage{rotating}

\slugcomment{Accepted by the Astrophysical Journal, Jan 24, 2013}

\shorttitle{Spitzer transits of GJ1214b}
\shortauthors{Fraine et al.}

\begin{document}


\title{Spitzer Transits of the Super-Earth GJ1214b and Implications for its Atmosphere}


\author{
Jonathan~D.~Fraine\altaffilmark{1}, Drake~Deming\altaffilmark{1}, 
Micha\"el~Gillon\altaffilmark{2}, Emmanu\"el Jehin\altaffilmark{2}, 
Brice-Olivier~Demory\altaffilmark{3}, Bjoern Benneke\altaffilmark{3}, Sara~Seager\altaffilmark{3}, 
Nikole~K.~Lewis\altaffilmark{4, 5, 6}, 
Heather~Knutson\altaffilmark{7}, \& 
Jean-Michel~D\'esert\altaffilmark{6, 8, 9}
}


\altaffiltext{1}{Department of Astronomy, University of Maryland, College Park, MD 20742 USA; jfraine@astro.umd.edu}
\altaffiltext{2}{Institute d'Astrophysique et de Geophysique, Universite de Liege, Liege, Belgium}
\altaffiltext{3}{Department of Earth, Atmospheric and Planetary Sciences, and Department of Physics,
     Massachusetts Institure of Technology, Cambridge, MA 02139 USA}
\altaffiltext{4}{Department of Planetary Sciences and Lunar and Planetary Laboratory, University of Arizona, 
  Tucson, AZ 85721 USA}
\altaffiltext{5}{Present address: Department of Earth, Atmospheric and Planetary Sciences, and Department of Physics,
  Massachusetts Institure of Technology, Cambridge, MA 02139 USA}
\altaffiltext{6}{Sagan Fellow}
\altaffiltext{7}{Division of Geological and Planetary Sciences, California Institute of Technology, 
  Pasadena, CA 91125 USA}
\altaffiltext{8}{Harvard-Smithsonian Center for Astrophysics, 60 Garden St., Cambridge, MA 02138 USA}
\altaffiltext{9}{Present address: Division of Geological and Planetary Sciences, California Institute of Technology, 
  Pasadena, CA 91125 USA}


\begin{abstract}
We observed the transiting super-Earth exoplanet GJ1214b using Warm Spitzer 
at 4.5\,$\mu$m wavelength during a 20-day quasi-continuous sequence in May 2011. 
The goals of our long observation were to accurately define the infrared 
transit radius of this nearby super-Earth, to search for the secondary eclipse, 
and to search for other transiting planets in the habitable zone of GJ1214. 
We here report results from the transit monitoring of GJ1214b, including a 
re-analysis of previous transit observations by \citet{desert11}. In total, 
we analyse 14 transits of GJ1214b at 4.5\,$\mu$m, 3~transits at 3.6\,$\mu$m, 
and 7~new ground-based transits in the I+z band. Our new Spitzer data by 
themselves eliminate cloudless solar composition atmospheres for GJ1214b, 
and methane-rich models from \citet{howe}. Using our new Spitzer 
measurements to anchor the observed transit radii of GJ1214b at long wavelengths, 
and adding new measurements in I+z, we evaluate models from \citet{benneke} 
and \citet{howe} using a $\chi^2$ analysis. We find that the best-fit model 
exhibits an increase in transit radius at short wavelengths due to Rayleigh 
scattering. Pure water atmospheres are also possible. However, a flat line 
(no atmosphere detected) remains among the best of the statistically acceptable 
models, and better than pure water atmospheres. We explore the effect of systematic 
differences among results from different observational groups, and we find that 
the \citet{howe} tholin-haze model remains the best fit, even when systematic differences 
among observers are considered.

\end{abstract}



\section{Introduction}

The mass and radius of the nearby transiting super-Earth GJ1214b
\citep{charb09} imply that it must have a significant atmosphere
\citep{rogers}.  That inference motivated an extensive effort to
detect the atmosphere, by seeking wavelength variations of the transit
depth.  A wide variety of compositions are possible for super-Earth
atmospheres \citep{miller-ricci, benneke, howe}, from
hydrogen-dominated to heavy-element-rich. Most current observations of
the transits \citep{bean10, bean11, crossfield, desert11, berta11,
berta12, deMooij} have rejected hydrogen-dominated atmospheres for
GJ1214b, but \citet{croll} concluded in favor of a low molecular
weight atmosphere.  The infrared (IR) spectral region is particularly
important for such studies, because strong water vapor bands increase
the transit depth in the IR significantly as compared to the optical.
This is especially true for hydrogen-dominated atmospheres, because of
the increased atmospheric scale height. The intrinsically strong IR
water vapor opacity makes hydrogen-dominated atmospheres opaque in the
IR over several scale heights, in spite of their relative paucity of
heavy elements.  Also, strong bands of methane and carbon monoxide fall
within the Warm Spitzer bandpasses at 3.6- and 4.5\,$\mu$m,
respectively. Considering that GJ1214b's M-dwarf host star is
bright in the IR, transit observations of GJ1214b using Warm Spitzer
become particularly relevant to the characterization of its
atmosphere.

In May 2011, we observed the GJ1214 system quasi-continuously for 20
days, using Spitzer's IRAC instrument at 4.5\,$\mu$m.  Our
investigation had three goals: 1) to improve the transit parameters of
the system and constrain the properties of the planet's transmission
spectrum, 2) to search for the secondary eclipse, and 3) to search for
other transiting planets in this system, to the outer edge of the
habitable zone.  

Our observations included minor interruptions for data downloads. But,
an unplanned 42-hour data loss also occurred during the 20-day
sequence, caused by a combination of spacecraft and Deep Space Network
(DSN) downlink anomalies.  We therefore re-observed GJ1214 for an
additional 42-hours in November, 2011, using the IRAC 3.6\,$\mu$m
band.  Consequently, we have multiple transits at both Warm Spitzer
wavelengths, and these data provide a particularly powerful constraint
on the IR transit depth.  Moreover, since our 4.5\,$\mu$m transits
were observed nearly consecutively, we have an excellent basis for
evaluating the degree to which stellar activity (e.g., star spots)
affect the inferred transit depth.

A global analysis of these GJ1214 data is being published by
\citet{gillon12}, including a search for other transiting planets in
this system.  In this paper, we focus on the implications of our
observations for understanding the nature of GJ1214b's atmosphere, and
we investigate the wavelength-dependent transit radius of GJ1214b in
detail.  Our analysis includes the degree to which star spots - even
those not occulted by the planet - contribute to possible bias in the
measured radius of the planet.  Anchored by our improved precision for
these infrared transits of GJ1214b, we add 7 new ground-based transits
in the I+z band, and we re-analyze the totality of published
wavelength-dependent transit depths for GJ1214b, exploiting recent
advances in super-Earth model atmospheres \citep{benneke, howe}.

Sec.~2 describes the details of our observations, and Sec.~3 explains
our procedure to extract precise photometry from the data.  In Sec.~4
we fit to the photometry, extracting the transit radius for the planet
in the warm Spitzer and I+z bands, and we derive improved system parameters.  
Sec.~5 considers the possible effect of star spots on our results.  Sec.~6
discusses implications for the atmosphere of GJ1214b.

\section{Observations}

\subsection{Spitzer}

We observed GJ1214 for 20 consecutive days using Warm Spitzer at
4.5\,$\mu$m, beginning on April 29, 2011 at 03-46 UTC.  We used
subarray mode with an exposure time of 2 seconds per frame.  The
observations contain several $\sim 3$-hour interruptions for data
download, and one unanticipated 42-hour gap where data were
irretrievably lost.  The data loss occurred because the DSN incurred
anomalous delays in downloading data at that time.  When the Spitzer
observatory was designed, long observing campaigns of exoplanet
photometry were not envisioned.  Spitzer's onboard flight software was
designed to automatically delete data after a certain period of time,
to make room in memory for new observations.  The DSN anomaly
consequently caused the onboard software to delete data before
downlink (the flight software has now been corrected so that this will
not recur).

To compensate for the data loss, we were awarded 42 hours of
continuous observations that began on November 6, 2011, at 11-54 UTC.  We elected
to acquire these data in Spitzer's 3.6\,$\mu$m bandpass, to complement
the 4.5\,$\mu$m data from May 2011.  We again used 2-second exposure
times in subarray mode.  In total, our data comprise 791,808 exposures
at 4.5\,$\mu$m, and 74,624 exposures at 3.6\,$\mu$m.

\subsection{TRAPPIST}

In order to help define the possible effects of stellar activity on
the Spitzer transits, to further cross-check our analysis versus
\citet{gillon12}, and to add additional information relevant to the
atmosphere of GJ1214b, we observed 7 transits using the TRAPPIST
facility \citep{gillon11, jehin}, over the period 2011 March 11 - May
18. The TRAPPIST observations and photometry are described by
\citet{gillon12}, but we summarize the data here.  The observations
were made using the 60-cm robotic telescope in a slightly defocused
mode. An I+z filter gave transmission from 750 to 1100 nm.
Differential photometry on the 25-sec exposure images was done (by
M.G.) using IRAF/DAOPHOT.  In our analysis, we use the same version of
the photometry as \citet{gillon12}, but we perform an independent
analysis and transit fitting.

\section{Spitzer Photometry}

\subsection{Aperture Photometry}

Our analysis utilizes the Basic Calibrated Data (BCD) files produced
by version S18.18.0 of the Spitzer pipeline.  Two dimensional (2D) 
Gaussian centering produces the least scatter in our final photometry 
\citep{stevenson, agol}.  \citet{knutson} found that flux-weighted 
centering gives superior results for Spitzer data in studies of 
exoplanetary phase curves over long time scales.  We tried flux-weighted 
centering for our transit analysis, but it did not result in significant 
improvement over our 2D Gaussian centering.

In the case of the 4.5\,$\mu$m photometry we center a circular
aperture of constant radius on the star.  We calculate the stellar flux
within the aperture, including analytic approximations for the
partial coverage of pixels at the boundary of the aperture.  We vary the
radius of the aperture from 2.0 to 5.0 pixels, in 0.5-pixel
increments, and thereby produce seven versions of the photometry at
each wavelength.  After decorrelation (see below), we chose to use an
aperture radius of 2.5 pixels for 4.5\,$\mu$m, based on the global
scatter in the decorrelated photometry.

In the case of the 3.6\,$\mu$m photometry, we examined constant radius photometry-- 
for the same seven radii as in the 4.5\,$\mu$m section-- and variable radius 
aperture photometry.  The variable radius photometry improves the precision by 
41\% over constant aperture photometry by varying the aperture radius as 
a function of the `noise-pixel'  calculations per frame \citep{mighell, knutson, lewis}. 
Therefore we adopt the  noise pixel method for our 3.6\,$\mu$m photometry, and 
subsequent intra-pixel decorrelation (see below).

The noise pixel method estimates the effective width of the pixel
response function, accounting for undersampling, by calculating the
variance of the flux per frame, weighted by the square of the mean
flux:

\begin{equation}
\tilde{\beta}_i = \frac{\left(\Sigma I_j\right)^2}{\Sigma \left(I_j\right)^2},
\end{equation}  

where $\tilde{\beta}_i$ is the noise pixel parameter for the stellar image in
frame $i$, $I_j$ is the intensity of pixel $j$, and the summations
extend over all pixels wherein the stellar intensity is significant.
Using the $\sqrt{\tilde{\beta}_i}$ as the aperture radius collects an optimum
amount of light for photometry.  The average aperture radius from
using this formulation on our 3.6\,$\mu$m data is 2.60 pixels.
\subsection{Decorrelation}

Upon producing photometry, we immediately see Spitzer's well-known
intra-pixel sensitivity effect, that must be decorrelated and removed
from the data. A portion of the raw 4.5\,$\mu$m photometry (before
decorrelation) is illustrated in Figure~1. Because the transits of
GJ1214b are substantially larger than the intra-pixel signature, we
first mask off the transits from the first stage of decorrelation.
Our 4.5\,$\mu$m data comprise seven distinct re-acquisitions of
GJ1214, with interruptions for data download. Therefore our first
stage of decorrelation is done separately for each of the seven
re-acquisitions.  That inital decorelation fits two polynomials to the
data for each re-acquisition. (There are seven sections of the data
defined by the data downloads and re-acquisitions.)  One polynomial
fit is applied to those data points wherein the image centroid lies at
Y-coordinate greater than the center of the pixel, and another
polynomial fit is applied to those data wherein the image lies at
Y-coordinate less than the center of the pixel. Our rationale for this
two-parameter fit is based on visual inspection of the photometry,
that shows different behavior above the center-of-pixel than below
center (see lower panel of Figure~1). The polynomials are fourth
order in Y and second order in X, because the photometric variations
as a function of Y are more pronounced than with X. Also, visual
examination of the variation with Y indicated that lower order
polynomials (e.g., quadratic) would not represent the variations
optimally.  After finding the best fit polynomials, we divide them
into the data, including the in-transit data.  We only used this step
to determine initial conditions for an iterative weighting function
method (see below).

We follow the polynomial decorrelation with a second procedure that
is itself a two-pass iterative process. We first remove a preliminary
transit model from the polynomial-decorrelated data, and apply a
Gaussian weighting function \citep{ballard, lewis, knutson} to correct 
the intra-pixel signatures.

The kernel of the Gaussian weighting uses a $\sigma$ of 0.005 pixels in Y, 
and 0.01 pixels in X for the 4.5\,$\mu$m data, that we determined by trial 
and error - evaluating the noise level of the final decorrelated photometry. 
A separate Gaussian weighting was applied to each section of the data between 
downloads, but the same kernel size was used for all data sections at 4.5\,$\mu$m.  

For the 3.6\,$\mu$m data, we applied a noise-pixel, Gaussian weighting 
function that uses a variable $\sigma$ in $Y$, $X$, and $\tilde{\beta}$ (noise-pixel value).  
This varied the weights in the Gaussian kernel to accommodate the 
necessary number of neighboring points that influence the strength of the 
correlation between center position and flux, as discussed in \citet{lewis} and \citet{knutson}.

Because of effects near the pixel boundary, we chose to subdivide the 3.6 $\mu$m 
data into 2 sections, encompassing the 2 transits in the data.  The pixel 
boundary effects occured well out of phase of both transits.  After examining 
the global scatter and $\sigma ~vs.~N^{-0.5}$ (bin size) slope residuals, we found 
that decorrelating after subdividing the 3.6 $\mu$m data set to within $\pm 0.05$ 
of each transit produced the best noise levels, and resulted in more conservative  
uncertainty estimates on the transit parameters, by $\sim 10$\%.  Moreover, 
this phase range coincides with similar subdivisions in the 4.5 $\mu$m data.  

After dividing by the results of the Gaussian weighting, we solve for
and subtract an improved transit model, and again apply a second stage 
weighting function decorrelation.  We experimented with a third iteration 
of this process, but it did not produce significant improvement. Unlike
the first (polynomial) stage of decorrelation, the Gaussian weighting
function was applied to the totality of the data (i.e., all
re-acquisitions) using a single Gaussian kernel. The initial
polynomial stage of the decorrelation process may seem unnecessary,
since the weighting function alone could remove structure in the data
on both large and small spatial scales.  However, we find that the
polynomials speed up the iterative process by providing a fast start
to the iteration. Our final photometry has a standard deviation of
$3.7 \times 10^{-3}$, which is only 15\% greater than the photon
noise.  Moreover, red noise is minimal and the precision for binned
data improves nearly as the square-root of the bin size, as we
demonstrate below.

Figure~2 shows an overview of our 4.5\,$\mu$m photometry, after
decorrelation; more detailed depictions are discussed and shown below.

\section{Model Fitting for Transit Parameters}

We analyze all available Sptizer transits, including a re-analysis 
of the transits reported by \citet{desert11}, and TRAPPIST transits 
using the I+z filter \citep{gillon12}.  We use three methodologies to 
determine the best-fit transit parameters for each dataset and wavelength. All 
three methods use only the data within a phase interval of $\pm0.05$ 
around the center of each transit-- except for the TRAPPIST transits, which 
used all of the available data from \citet{gillon12}. 

The first method solves for the best-fit transit parameters of all
transits simultaneously at each wavelength.  The second method fits
each transit individually and independently, then calculates the
average of the transit parameters at each wavelength, weighting the
individual results by the inverse of their variance. The third method
phases and bins all of the transits at each wavelength into a single
transit, and fits to those phased \& binned data.  All 3 methods
included a total of 3 transits at 3.6\,$\mu$m, 14 transits at 4.5\,$\mu$m, 
and 7 transits in the I+z band. Comparing these three methods gives an
indication of the consistency of our results, and we do indeed find
good consistency, as noted below.

We now describe the details of those fitting procedures.

\subsection{Spitzer 4.5\,$\mu$m Transits}

We use the formulation of \citet{mandel} to generate transit curves
and fit them to the observed data, thereby extracting the essential
parameters of the transit.  The 14 Spitzer transits at 4.5\,$\mu$m
comprise our most extensive and highest quality data. Our simultaneous
fit holds the orbital period fixed at the value measured by
\citet{bean11} $\left(1.58040481 \pm 1.210^{-7}\right)$, and uses a 
Levenberg-Marquardt algorithm to minimize the $\chi^2$ for each transit.  
Although we are minimizing $\chi^2$, we do not accept that particular 
set of transit parameters as our best-fit values.  Instead, we explore 
parameter space using a Markov Chain Monte Carlo (MCMC) method, and also 
using a residual-permutation (`prayer-bead') method.  Our best-fit values 
are taken from the medians of the posterior distributions generated in 
this exploration of parameter space (see Sec.~4.4).

Our fit extracts a correction to the transit center time (as might be
caused by ephemeris error), as well as $a/R_{*}$, $i$, $R_p/R_{*}$,
and a linear limb-darkening coefficient.  We hold the quadratic limb
darkening coefficient at zero because Bayesian Information Criterion
\citep{liddle} analysis supports a linear law, which gives an adequate 
account of the minimal limb darkening that is characteristic of infrared 
transits. Moreover, our derived linear limb darkening coefficients at 3.6 
and 4.5 microns (c0 =0.158 and 0.128 respectively, see Table 4) are 
reasonably consistent with the values predicted by \citet{claret} (c0= 0.147 
and 0.155) for a model atmosphere having T=3500K (that temperature being 
their closest match to GJ1214).  

Figure~3 zooms in on a portion of the simultaneous fit for the 
first two 4.5\,$\mu$m transits, and Figure~4 shows the standard deviation 
of residuals (data minus simultaneous fit) when binned over different
time intervals.  Note that we find very little red noise in these
data, indicating the success of our multi-stage intra-pixel
decorrelation.  Note also that our decorrelation process will tend to
remove slow variations in the stellar brightness.  Hence the
apparently constant flux seen in Figure~2 should not be interpreted as
evidence for stellar quiesence.  (The possible effects of stellar
activity on our results are discussed in Sec.~5).

In addition, we bin all 14 4.5 $\mu$m transits, including that of
\citet{desert11}, into bins of width $0.001$ in phase, using a running standard 
deviation for the weights in a weighted mean-- phasing them all to a common 
epoch determined from the simultaneous fit.  As before, we include only data 
within a phase range $\pm0.05$ of transit center, and we fit to the phased \& binned transit 
using the same Levenberg-Marquardt algorithm described above.  Figure~5 shows the 
resulting fit in comparison to the phased \& binned data. 

The best-fit Spitzer transit parameters for each 4.5\,$\mu$m transit
fitted individually are given in Table~1; the individual 3.6\,$\mu$m
transits (Sec.~4.2) are given in Table~2, and the individual TRAPPIST
results (Sec.~4.3) are given in Table~3.  Results using the
combining methods are summarized in Sec.~4.5.

\subsection{Spitzer 3.6\,$\mu$m Transits}

Our 42-hour `replacement' observations contain two transits at
3.6\,$\mu$m, one near the beginning of these data and one near the
end.  The photometry for these transits was decorrelated using the
same methodology described above for 4.5\,$\mu$m. In contrast to the
4.5\,$\mu$m case, our 3.6\,$\mu$m photometry exhibits noticeable red
noise. Fortunately, this red noise is most significant in the long
interval between the two transits.  We limited the effect of this red
noise by limiting the range of the data included in our decorrelations 
and fits.  The omitted data did not occur near the transits, nevertheless 
we tried to develop objective criteria for the range of data that were used.  

After comparing the $\sigma ~vs.~N^{-0.5}$ and global $\sigma$, or RMS scatter, 
of various sized data slices, we determined that trimming the 3.6 $\mu$m data set 
into a phase range of $\pm 0.05$ around the center of each transit minimized the 
RMS scatter and the residual of the $\sigma ~vs.~N^{-0.5}$ slope, which minimized 
the red noise for these transits; and was fortuitously the same range as the $\pm0.05$ 
in phase that we adopted for our 4.5\,$\mu$m fits.

We repeated all of the methodology described in this section on the 
\citet{desert11} 3.6\,$\mu$m transit data, and we found that the slope of 
$\sigma~vs.~N^{-0.5}$ was insensitive to the range of data analyzed.  As a 
result, we use the entire \citet{desert11} 3.6\,$\mu$m data set.

As in the 4.5\,$\mu$m case, we fit to all 3 of the 3.6\,$\mu$m transits 
using 3 methods: simultaneously, individually, and phased \& binned.  
Figure~6 shows all 3 of the 3.6\,$\mu$m transits phased \& binned, overlaid 
with a best-fit curve.  Figure~7 illustrates the standard deviation of the 
3.6\,$\mu$m residuals as a function of bin size for our fits to the simultaneous 
transit parameters.

\subsection{TRAPPIST I+z Transits}

For the TRAPPIST data set, we used all 7 distinct epochs of the  GJ1214b 
transiting system provided by \citet{gillon12}.   Three of these epochs 
overlapped with the Spitzer 4.5\,$\mu$m data set.  \citet{gillon12} used 
the I+z filter on the TRAPPIST telescope because it supplied a near uniform
filter profile from 0.7 - 1.0 $\mu$m.  

We determined the physical parameters using all three methods discussed above: a 
simultanous fit, individual fits, and a fit to phased \& binned data.  Similar to 
the Spitzer data sets, we fit a \citet{mandel} transit model using a Levenberg-Marquardt 
routine.  In contrast to the Spitzer data sets, we fit the TRAPPIST data using a 
\citet{mandel} model that included quadratic limb darkening, which accounted for 
the excess stellar limb-darkening observed in the shorter wavelength transit data.
The fit to the phased \& binned TRAPPIST data is shown as Figure~8.  To analyze the 
quality of the fit, Figure~9 shows the $\sigma ~vs.~N^{-0.5}$ for the simultaneous 
fit of the TRAPPIST data set.

The results for the planet-to-star radius ratios, and related error bars, in 
the I+z band are plotted on Figure~10 for comparison to our 13 Spitzer transits 
at 4.5\,$\mu$m, and listed in Tables~3 \& 4.  Further information about the data 
reduction process for the TRAPPIST data set is included in \citet{gillon12}.

\subsection{Errors}

We used two methods to estimate uncertainties for our derived transit
parameters: MCMC and prayer-bead.  In both methods the errors - as
well as the best-fit values - follow from the posterior distributions.
We adopted the prayer-bead method for our quoted results because it
explicitly includes the effect of red noise \citep{desert11b}.
Figure~11 shows the distributions for $R_p/R_*$ for one of our
4.5\,$\mu$m individual fits, showing a broader distribution for the
prayer-bead method.

In implementing the prayer-bead method, we permute the residuals only
within the adopted phase range of the fit ($\pm0.05$ in phase).  At
each permutation, we find the best-fit transit parameters using the
Levenberg-Marquardt method, and we add those best-fit values to the
posterior distributions of each parameter. We adopt the median of the
posterior distribution as the best fit value, following
\citet{desert11b}.


A potential additional source of error is associated with the
decorrelation, that is not explicitly propagated into the stage of
fitting the photometry.  However, we are not concerned about this for
two reasons.  First, the 4.5\,$\mu$m data are so extensive, and the
intra-pixel effect is so modest at that wavelength, that we believe
those errors in decorrelation have negligible effect.  Second, our
procedure accounts for imperfect decorrelation at both Spitzer
wavelengths, in an {\it implicit} fashion.  Imperfections in decorrelation 
create red noise, and that red noise contributes to errors on the derived
transit parameters using the prayer-bead method.

In addition to random error, systematic differences may exist between
our results and other investigators.  We discuss one source of
possible systematic difference in Sec.~4.5; the implications of these
differences for the nature of GJ1214b's atmosphere is discussed in
Sec.~6.

\subsection{System Parameters}

As noted above, we estimated the system parameters using three
methods: 1) simultaneous fitting of all transits at a given
wavelength, 2) averaging system parameters from the individual fits to
each transit at a given wavelength, and 3) fitting to phased \&
binned combinations of transits at a given wavelength.  We included
the transits observed by \citet{desert11} in all three methods.  We 
also fit the TRAPPIST transits using all three methods (see Table~4).
Table~1 lists the individual fits to the 4.5\,$\mu$m data; Table~2
gives the individual fit results at 3.6\,$\mu$m and Table~3 lists the
individual fit results for the TRAPPIST transits in the I+z band.

Since our different fitting methods (summarized in Table~4) are simply
different ways of accounting for the same data, they should give
consistent results as far as the best-fit parameters are concerned.
Beyond best-fit consistency, we find that comparison of these methods
can provide a basis for caution concerning the errors on the derived
parameters. For example, our 4.5\,$\mu$m results from the simultaneous
fit give ${R_p}/{R_*} = 0.11710\pm0.00017$, whereas averaging the
individual fits, weighted by the inverse of their variances, gives
${R_p}/{R_*} = 0.11699 \pm 0.00026$.  Although the best-fit values
agree well, the larger error from averaging the individual fits may
indicate potential variations from transit-to-transit.

Table~5 summarizes our results for the ${R_p}/{R_*}$ parameter that
potentially reveals information about the atmosphere of GJ1214b into one
concise Table.  For the discussion that follows, we adopt the results
from the phased \& binned method, because we feel that the high
precision of these combined transit curves allows the most reliable
solution.

We explored further comparison with two other precise measurements of
$R_p/R_{*}$: \citet{bean11} and \citet{berta12}.  We choose these
investigations for more in-depth comparison because the former is the
highest precision ground-based measurement of GJ1214b, and the latter
is a precise space-borne measurement.  We investigated to what extent
differences in $R_p/R_{*}$ arise from the different transit solutions,
with different values for orbital transit parameters such as $a/R_{*}$
and orbital inclination.  Because these orbital parameters should not vary 
with wavelength, we force our solutions to adopt the values as derived by 
\citet{bean11} and \citet{berta12}. Our resultant retrievals for 
$R_p/R_{*}$ are included in Table~5.  On average, constraining our orbital 
parameters to have the values found by either \citet{bean11} or 
\citet{berta12} results in decreasing our $R_p/R_{*}$  value at 4.5\,$\mu$m 
by about $0.0010$, but with less difference at 3.6\,$\mu$m or I+z. 

Arguably, we should adopt these constrained values as our principal result.  
However, the low limb darkening that prevails at Spitzer wavelengths, in 
combination with the high precision we achieve from our large dataset, 
motivates us to rely primarily on our own orbital parameters.  Nevertheless, 
we explore the implications of adopting the \citet{bean11} and \citet{berta12} 
orbital parameters in Sec.~6.

\section{Transit-to-Transit Variability and Star Spots}

A star spot crossing during transit appears as an anomalous spike or
bump in the transit light curve (e.g., \citealp{deming11}). Our photometry
shows no evidence that the planet crossed even one significant star
spot during our thirteen 4.5\,$\mu$m transits. Nevertheless, spots are
common on M-dwarf stars, and uncrossed star spots could still affect
the transit \citep{sing, desert11b}.

TRAPPIST photometery of GJ1214 out of transit, but over the same time
period as our 4.5\,$\mu$m Spitzer observations, shows essentially no
variation in the I-band \citep{gillon12}, to a limit of about 0.2\%.
Nevertheless, other investigations have found that GJ1214 exhibits
rotationally-modulated signatures of star spots, so we consider the
potential impact of such variation on our results.  As we will
demonstrate below, even allowing for more photometric variation
than \citet{gillon12} observed, star spots have negligible effect on
our results.

\citet{berta11} found that GJ1214 shows a photometric variation of 1\%
amplitude (2\% peak-to-peak) in the I-band (0.715-1.0\,$\mu$m) with a 
rotation period of $\sim\,53$ days, based on MEarth data \citep{nutzman}, 
spanning three years of observations.  Using PHOENIX model atmospheres
\citep{allard}, we calculate that the \citet{berta11} amplitude of
variation could be produced by two star spots, each covering as much
as 2\% of the sky-projected stellar disk, separated in longitude by
180$^\circ$. Based on Doppler imaging studies of active dwarf stars
\citep{rice}, we adopt a temperature contrast (spot vs. photosphere)
$\Delta T/T = 0.1$, thus $T_{spot} \sim 2700K$.  Note that this is the
same as adopted by \citet{berta11}. We use limb darkening coefficients
for both the photosphere and the star spot, calculated from our
transit fitting at 4.5\,$\mu$m ($c_0 = 0.11$, $c_1 = 0.0$). On this basis,
we determined that a 1\% variation of the stellar light curve in the
I-band translates to 0.42\% variation in Spitzer photometry at
4.5\,$\mu$m.

We developed a numerical tile-the-star model to calculate the effect
of unocculted star spots on the transit depth.  We created a synthetic 
time-series of 2D images of GJ1214, at 4.5\,$\mu$m, and projected two
circular star spots on its surface at the equator, separated by
180$^\circ$ of longitude.  This arrangement of spots in opposite
hemispheres produces an appropriate quasi-sinusoidal effect in the
total stellar light. 

We accounted for variation in the spots projected area as the star 
rotates, but we ignored the Wilson depression effect. The PHOENIX model 
with $T = 3000K$, $[M/H] = +0.3$, $log(g) = 5.0$ and $\alpha = 0.0$ 
represented GJ1214; and, the PHOENIX model with $T = 2700K$, $[M/H] = 
+0.3$, $log(g) = 5.0$, $\alpha = 0.0$ represented the star spot.  We 
multiplied Spitzer's 4.5\,$\mu$m filter profile by the spectral models 
and integrated over wavelength to calculate the expected 4.5\,$\mu$m 
flux variations due to stellar rotation with the spots fixed in longitude.  
Both the star and spots are affected by limb darkening as determined by 
our fitted \citet{mandel} model parameters.

It is easy to show that the effect of unocculted star spots on the
planetary radius derived from the transit is given as:

\begin{equation}
\left(\frac{R_p}{R_*}\right)^2_{spotted} 
= \frac{F_{ot} - F_{it}}{F_{ot}}
= \frac{\left(\frac{R_p}{R_*}\right)^2_{spotless} I_{ph}}{(1-\epsilon) I_{ph} + \epsilon I_{spot}}
\end{equation}  

\noindent with $\epsilon = \frac{A_{spot}}{\pi R_*^2}$, and where $ot$
indexes out of transit, $it$ indexes in transit. $I$ indicates the
intensity of the stellar disk, $I_{ph}$ being the intensity of the
photosphere, and $I_{spot}$ being the intensity of the spot.

Using this equation together with flux variations from our
tile-the-star model, we calculated the potential variation in transit 
depth as a function of time.  This quasi-sinusoidal variation has an unknown
phase because we do not know the longitudes of any real star spots on
GJ1214.  Nevertheless, the amplitude of variation in $R_p/R_*$ from
this model is $9.6 \times 10^{-5}$, which is negligible compared to
the observed scatter in our measurements (see Figure~11).  Moreover, as
noted above, the photometric variations of GJ1214, observed
concurrently with our transit data \citep{gillon12}, were much less
than from \citet{berta11}. We therefore conclude that star spots play
a negligible role on the observed variations and/or possible bias of
our inferred radii for GJ1214b.  However, we cannot exclude the possibility 
that increased star spots during \citet{bean11} and \citet{berta12} observations 
are responsible for some of the differences between our results. 

\section{The Atmosphere of GJ1214b}

There are numerous transit observations in the literature that bear 
on the nature of the atmosphere of GJ1214b \citep{charb09, bean10,
bean11, desert11, berta11, berta12, croll, sada, carter, kundurthy, 
deMooij, murgas, narita}.  Nevertheless, the improved precision we 
have been able to achieve in the Spitzer bands, together with the new 
TRAPPIST results in the I+z-band, and recent advances in modeling 
the atmosphere of GJ1214b \citep{benneke, howe}, motivate us to re-
examine the nature of GJ1214b's atmosphere.  

In the following sub-sections we review the methodology for comparing 
observations and models (Sec.~6.1), we briefly preview what our Spitzer 
observations alone can reveal concerning the atmosphere of GJ1214b 
(Sec.~6.2), and we then compare the totality of all published 
observations  to existing models, using a $\chi^2$ analysis (Sec.~6.3).  
Since planetary radii derived by different observational groups can 
differ systematically, we discuss the effect of one particular 
systematic difference in Sec.~6.4, and we summarize our conclusions 
concerning the atmosphere of GJ1214b in Sec.~6.5.

\subsection{Comparing Models to Observations}

To compare transmission models for GJ1214b with the observations, we
need to integrate the models-- multiplied by the filter profiles--  over the 
observed bandpasses.  Let  $F_{ot}(\lambda)$ be the out of transit flux 
measured from the star as a function of wavelength.  Similarly let 
$F_{it}(\lambda)$ be the in-transit flux as a function of wavelength. Consider 
the simplified case where limb darkening can be neglected (arguably applicable 
in the infrared),  and include the fact that there is a wavelength-dependent 
observational sensitivity, $S(\lambda)$.  In that case:

\begin{equation}
F_{it}(\lambda) = S(\lambda) I_{*}(\lambda) (\pi R_{*}^{2} - \pi R_{p}(\lambda)^{2}),
\end{equation}

where $I_{*}(\lambda)$ is the intensity emergent from the stellar
atmosphere at wavelength $\lambda$.  The out of transit flux is:

\begin{equation}
F_{ot}(\lambda) = S(\lambda) I_{*}(\lambda) \pi R_{*}^{2}.
\end{equation}

With realistic spectral resolution, the observed quantities are integrals over
the observational bandpass, and the transit depth $d$ is:

\begin{equation}
d = \int [{F_{ot}(\lambda) - F_{it}(\lambda)}]d\lambda / \int{F_{ot}(\lambda)} d\lambda,
\end{equation}

Thus:

\begin{equation}
d = \int S(\lambda)I_{*}(\lambda)R_{p}(\lambda)^{2} d\lambda / \int S(\lambda)I_{*}(\lambda) R_{*}^2 d\lambda.
\end{equation}

When we seek to evaluate a model of the planet's transit radius as a
function of wavelength ($R_{p}(\lambda)$), it is necessary to include
the wavelength dependence of the stellar intensity as well as the
observational sensitivity.  The latter is commonly incorporated in
numerous studies of both transits and secondary eclipses, but the
necessity of including the stellar intensity is less widely
appreciated, especially the possible effect of line structure in the
stellar spectrum. For example, if the stellar spectrum contains water
vapor absorption that overlaps to some degree with planetary water
vapor features, then the apparent transit depth will be reduced
compared to the case where the star is purely a continuum source.
Stellar intensity weighting is particularly important for M-dwarf host
stars like GJ1214, because their spectrum varies strongly with
wavelength in the optical and near-IR.  Unfortunately, that weighting
is also considerably uncertain for M-dwarf stars, particularly at the
very interesting blue wavelengths where the planet may exhibit 
scattering from haze \citep{benneke, howe}.

To compare models of $R_{p}(\lambda)/R_{*}$ to observations, we
calculate the value of $d$ using Eq. (6), and infer $R_{p}/R_{*}$ as
$\sqrt d$.  This procedure is valid even at wavelengths where
appreciable limb darkening prevails. 


Although we use a PHOENIX model atmosphere to perform the weighting
over the observed bandpass, this model is not ultimately satisfactory
for this purpose.  For example, in the green bandpass ($0.46\;\mu$m) where 
a transit was observed by \citet{deMooij}, the PHOENIX model has essentially 
no flux (many orders of magnitude below the peak flux), whereas the real
star has sufficient flux to produce a transit having good
signal-to-noise \citep{deMooij}. The reason is that the model includes
only LTE thermal emission from the star, and does not incorporate the
various emission signatures of magnetic activity.  Therefore we can
use the PHOENIX model only in the red-optical and infrared.  Our
default procedure is to hold the stellar intensity constant for
wavelengths shortward of 1000 nm (i.e., we set $I_{*}(\lambda) =
I_{*}(1000)$ for $\lambda \leq 1000$\,nm), but we verified that using
other prescriptions shortward of this limit (e.g., blackbody spectra)
do not greatly influence our present results.  However, as the
precision of observations improves, it will eventually be necessary to
have an accurate spectrum for the host star at all wavelengths.

\subsection{Implications from Spitzer} 

Prior to an exhaustive analysis of all data versus all models, we
mention what our new Spitzer data alone immediately reveal and/or
constrain concerning the atmosphere of GJ1214b.  We (BB \& SS)
generated three new model atmospheres for GJ1214b at an equilibrium
temperature of 546 K, based on the methodology of \citet{benneke}. The
models are: 1) a H-rich solar abundance model, 2) a `Hot-Halley'
composition model which begins with solar composition and adds minor
molecular constituents from accreted icy material \citep{benneke}, and
3) a pure water vapor atmosphere.  Figure~12 shows the result of
integrating these three models over the Spitzer observational
bandpasses, and including the stellar intensity using a PHOENIX model
having Teff/log(g)/[M/H] = 3000 K/5.0/0.3. Based on this comparison, the
solar composition model is eliminated based on the Spitzer data alone.
The water vapor model is preferred over the hot Halley model, based on
the $\chi^2$ analysis described below.  Moreover, we expect that
methane-rich models having large scale heights (not illustrated) will
be rejected by the Spitzer data, due to the relative lack of an
enhanced radius in the 3.6\,$\mu$m band - that contains the strong
$\nu_3$ band of methane.

\subsection{A $\chi^2$ Analysis}

To gain further quantitative insight, we weighted the three models
from \citet{benneke} discussed above, as well as all of the models
given by \citet{howe} over the observed bandpasses of all extant
transit observations of GJ1214, including the TRAPPIST and warm Spitzer
data reported here.  We fit the models to the data using a $\chi^2$
analysis, as described below. \citet{berta12} applied a similar
$\chi^2$ analysis to data from Hubble/WFC3, but not to the complete set 
of data that we do here, and in particular not including our new and
precise results in the warm Spitzer bands.  The utility of $\chi^2$ is 
well known to be problematic when combining data from different
observational groups.  However, $\chi^2$ is at least an objective way
to compare different models, and we explicitly consider one cause of
observer-to-observer systematic differences in Sec.~6.4.

To evaluate and compare possible models of the planetary atmosphere,
we fit each model to the data by adding an adjustable constant (i.e.,
wavelength-independent offset) to the modeled values of $R_{p}/R_{*}$.
We choose the constant to minimize the $\chi^2$ of the difference
between the adjusted model and the data.  Adding this constant is
equivalent to increasing (or decreasing) the size of the opaque (i.e.,
solid) portion of the planet by a small amount.  That effectively
varies the surface gravity of the planet, and would strictly speaking
be inconsistent with the model that is being adjusted.  However, the
planetary radius at a given atmospheric pressure level is not known at
a level of accuracy comparable to the adjustments we are making.
Moreover, the requisite adjustments in the model output (typically,
$0.0005$ in $R_{p}/R_{*}$), correspond to less than 1\% differences in
surface gravity.  We therefore find this procedure to be a valid and
useful tool for testing models versus the observations, and we note
that a similar procedure was used by \citet{berta12}.  We fit the
three models shown in Figure~12 \citep{benneke}, together with all of
the models from \citet{howe}, and we calculate $\chi^2$ values for
each fit. We also fit a flat line to the data, i.e., a planetary
radius that does not vary with wavelength - indicating no signature of
the atmosphere.

Several of the best fitting models are shown in Figure~13, compared to
the entirety of published radii for this planet.  In total, there are
97 observations of $R_{p}/R_{*}$ versus wavelength on Figure~13. The
water atmosphere from Figure~12, and two models from \citet{howe}, are
also included on Figure~13.  For the Figure, the models are
overplotted monochromatically, without integrating over the
bandpass. However, the $\chi^2$ values are calculated by integrating
the model over the bandpass of each observation as described above,
and adopting the observed errors from each source. In the case of
multi-band analyses \citep{kundurthy}, we use the total bandpass from
multiple filters.  For the three overplotted models we also show the
values for the integrals over the Spitzer bandpasses, as open symbols.

With 96 degrees of freedom, models can only be rejected at the 99.9\%
confidence level if they have a $\chi^2$ exceeding 144.6.  Among the
possible models, The pure water atmosphere from \citet{benneke} yields
$\chi^2=142.7$ for 96 degrees of freedom.  The hot Halley and solar
composition models from \cite{benneke} have $\chi^2$ values of 167.9
and 1054.7 respectively, and are unlikely descriptions of GJ1214b's
atmosphere.  The solar composition model in particular is strongly
rejected unless high clouds are included (see below). In this respect,
we note that some discussion of a solar composition atmosphere has
occurred with respect to transit observations near 2\,$\mu$m
\citep{croll, bean11}.  We here emphasize that a cloudless solar
composition model is incompatible with the Spitzer data alone, as well
as with the totality of the observations over all wavelengths. The
issue of the transit depth near 2\,$\mu$m - while an important datum -
is not crucial to rejecting a solar composition atmosphere.

Our improvement in the observed Spitzer precision at 3.6\,$\mu$m -
overlapping the strong $\nu_{3}$ band of methane, prompts us to
investigate the methane-composition models of \citet{howe}. All of
their methane compositions have $\chi^2$ above 184.6 (the value for 1\%
methane). However, a solar composition model having high clouds (down to
pressures of 0.1 mbars) produces an acceptable $\chi^2=145.5$. This model 
is plotted as  the red  line on Figure~13; it predicts an increase in 
radius in the difficult-to-observe regions near 2.7 and 3.3\,$\mu$m.  The 
best fit to the data is the yellow line on Figure~13, having $\chi^2=119.8$.  
This model, from \citet{howe}, contains a dense haze of small (0.1\,$\mu$m) 
tholin particles, extending to very high altitudes (1\,$\mu$bar).  It was 
disfavored by \citet{howe}, but our methodology is different in that we 
incorporate a marginally additive constant when fitting the models.  However, 
we point out that also among the acceptable models is a flat line (not 
illustrated on Figure~13).  This null hypothesis (no atmosphere detected) 
yields $\chi^2=137.0$.

\subsection{Systematic Differences Between Observers}

One further check on the acceptable models is to focus on the
difference between our precise Spitzer radii and very precise radii 
derived at other wavelengths, i.e., by \citet{bean11} and \citet{berta12}.  
As noted at the end of Sec.~4.5, we fix the variable orbital parameters
($a/R_{*}$ and $i$) at the values derived by \citet{bean11} and
\citet{berta12} and derive a lower value of $R_p/R_{*}$ at 4.5\,$\mu$m by 
about 0.001 (Table~5).  Using that alternative value in our $\chi^2$ analysis is
one way of evaluating the possible effect of observer-to-observer
differences in radii. This procedure increases the $\chi^2$ values of
most models, but by varying amounts.  Interestingly, the two least
affected models are the flat line, whose $\chi^2$ increases by
$\delta\chi^2=+1.6$, and the tholin-haze model.  The water model from 
\citet{benneke} has $\delta\chi^2 = +8.4$, and their hot Halley model 
has $\delta\chi^2 =+19.8$.  The scattering tholin-haze model from 
\citet{howe} remains as the lowest absolute $\chi^2$, and has 
$\delta\chi^2 = -2.3$, i.e. it becomes more likely, not less likely.  We 
conclude that, even when systematic differences among observational groups 
are considered, a scattering atmosphere is currently the best estimate for 
GJ1214b. However, we are unable to explore the vast phase space of other 
possible sources of systematic error. Hence we also conclude that the null 
hypothesis (no atmosphere detected) remains among the most favored models, 
especially when systematic errors are considered.

\subsection{Summary of Implications for the Atmosphere of GJ1214b}

We have obtained new radii for GJ1214b in the I+z band using TRAPPIST,
and very precise radii at 3.6- and 4.5\,$\mu$m using Warm Spitzer in a
long series of new observations.  Our $\chi^2$ analysis indicates that
the best-fit model for the atmosphere of GJ1214b contains a haze of
small particles extending to high altitudes, although pure water vapor
models remain a possibility.  However, a flat line is among the
best-fitting models, particularly when observer-to-observer systematic
differences are considered.  Therefore we extend the conclusion of
\citet{berta12} concerning the flatness of the transmisison spectrum
from 1.1- to 1.7\,$\mu$m to include our new high-precision Spitzer
measurements at 3.6- and 4.5\,$\mu$m.  The atmosphere of GJ1214b is not
unequivocally detected at this point in time.
\newline\newline\newline\indent
This work is based on observations made with the Spitzer Space Telescope, 
which is operated by the Jet Propulsion Laboratory, California Institute 
of Technology under a contract with NASA. Support for this work was 
provided by NASA through an award issued by JPL/Caltech. We thank the 
Spitzer staff for their hard work and dedication in implementing these 
difficult observations, and the anonymous referee for a careful review 
of this paper.

\clearpage


\begin{figure}
\epsscale{.85}
\plotone{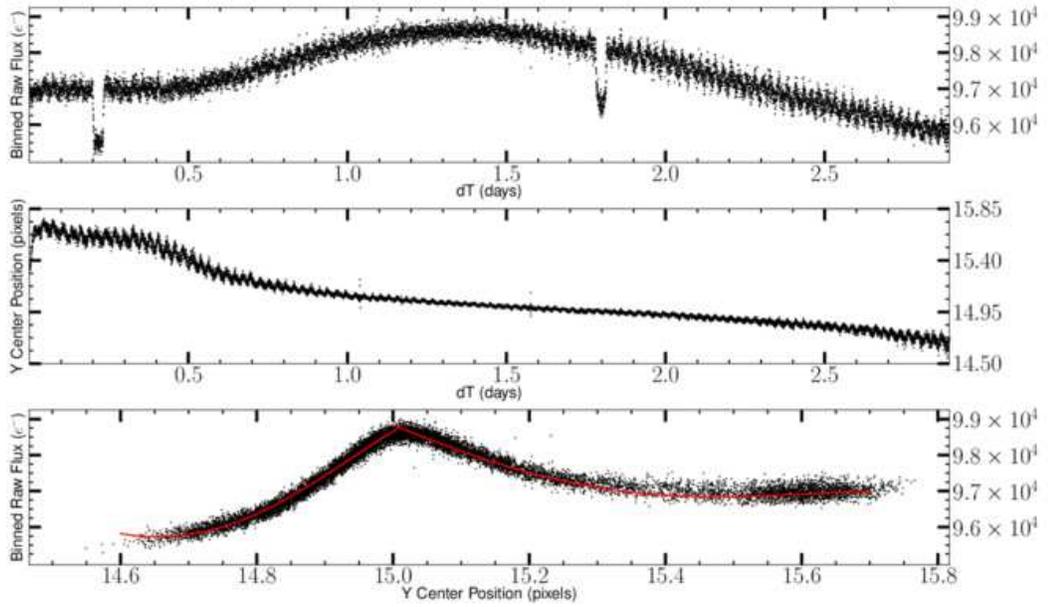}
\vspace{0.7in}
\caption{The top panel illustrates raw data for the first 2.9-day portion of
our 4.5\,$\mu$m photometry, containing two transits.  Each data point is aperture
photometry before decorrelation, binned by every 10 data points.  The middle 
panel shows the Y-center position as a function time; that which we correlate with
the flux to remove the intra-pixel effect, and related systematic noise sources.
The lower panel shows the same photometric values, except with the transit regions 
excluded, plotted versus the Y-pixel position of the stellar image.  Note the different 
spatial dependence of the photometry on each side of pixel center.  The red curves are
the 4th order polynomial fits that we use to initiate the decorrelation process (see text).
\label{fig1}}
\end{figure}

\clearpage

\begin{figure}
\epsscale{.85}
\plotone{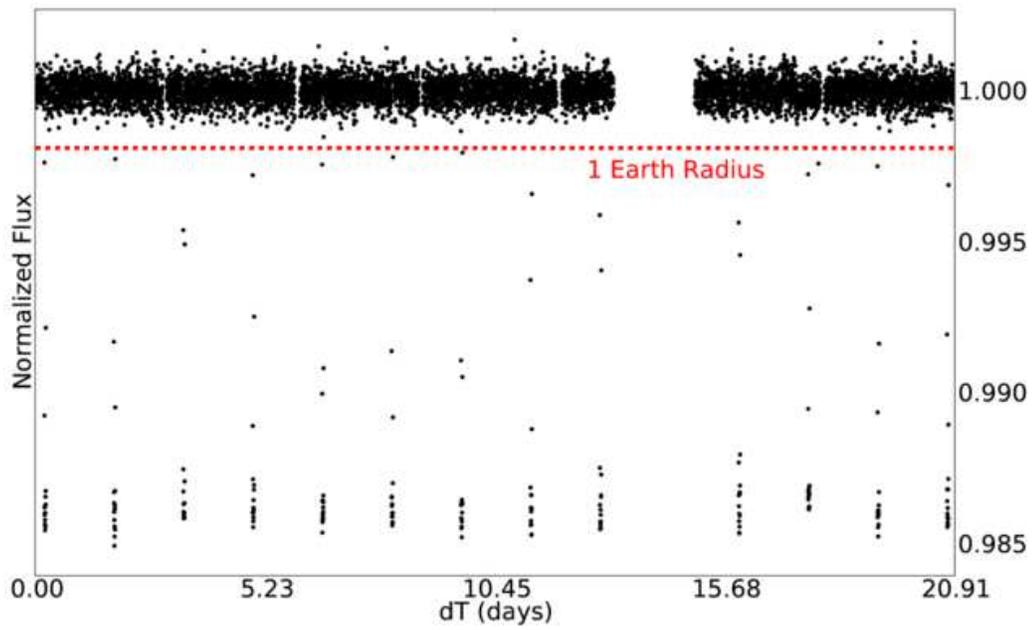}
\vspace{0.7in}
\caption{Overview of our 4.5\,$\mu$m photometry, after decorrelation
and binned in 100 two-second exposures per plotted point.  The dashed
line shows the transit depth that corresponds to one Earth radius. The
13 transits of GJ1214b are apparent. See \citet{gillon12} for an
analysis of other possible transiting planets in this system.
\label{fig2}}
\end{figure}

\clearpage

\begin{figure}
\epsscale{.70}
\plotone{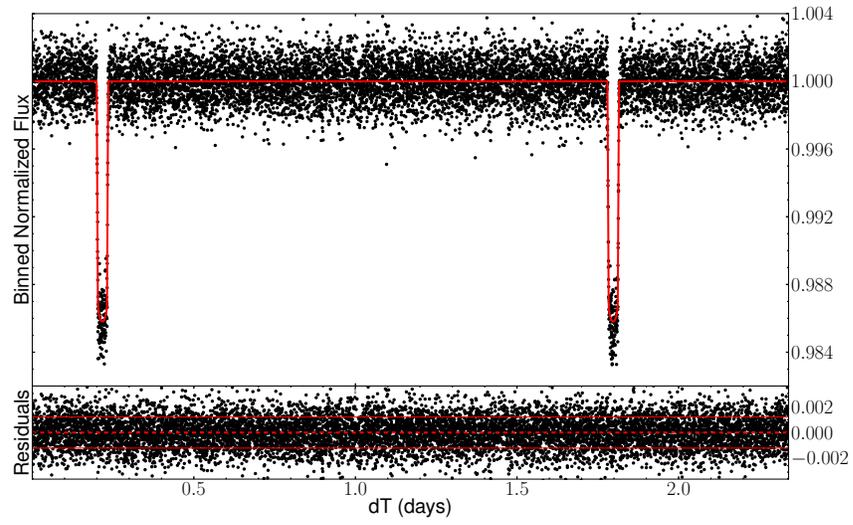}
\vspace{0.7in}
\caption{Expansion of a two-transit portion of the 4.5\,$\mu$m
photometry (Figure~2), with the best-fit transit curves overlaid, binned by every 10 data points.  
The red lines in the lower panel show a $\pm1\sigma$ envelope. 
\label{fig3}}
\end{figure}

\clearpage
		   
\begin{figure}
\epsscale{.85}
\plotone{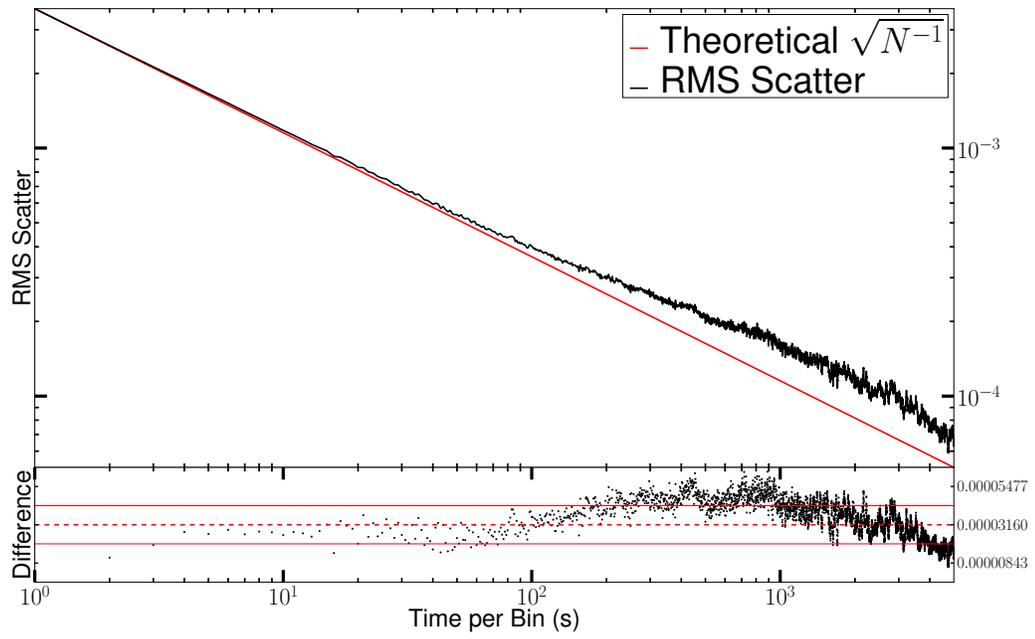}
\vspace{0.7in}
\caption{Standard deviation of the residuals (data minus the `simultaneous' fit, in units of
the stellar flux) for all 14 transits at 4.5\,$\mu$m, versus bin size. The red lines in the lower
panel show a $\pm1\sigma$ envelope. 
\label{fig4}}
\end{figure}

\clearpage
		   
\begin{figure}
\epsscale{.85}
\plotone{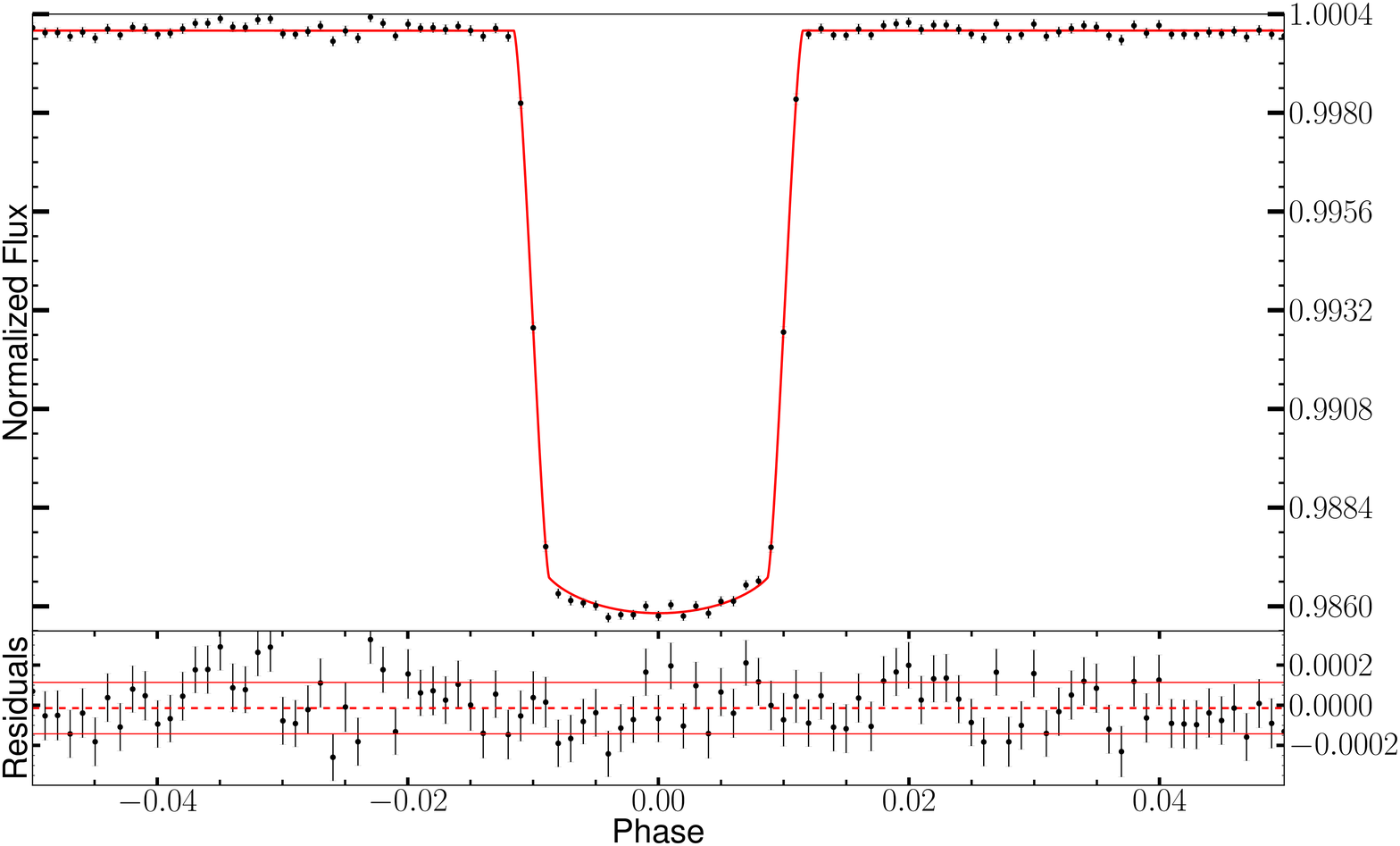}
\vspace{0.7in}
\caption{Phased \& binned transit of GJ1214b at 4.5\,$\mu$m from our 13 Spitzer transits, plus the transit observed by \citet{desert11}.
\label{fig5}}
\end{figure}

\clearpage

\begin{figure}
\epsscale{.85}
\plotone{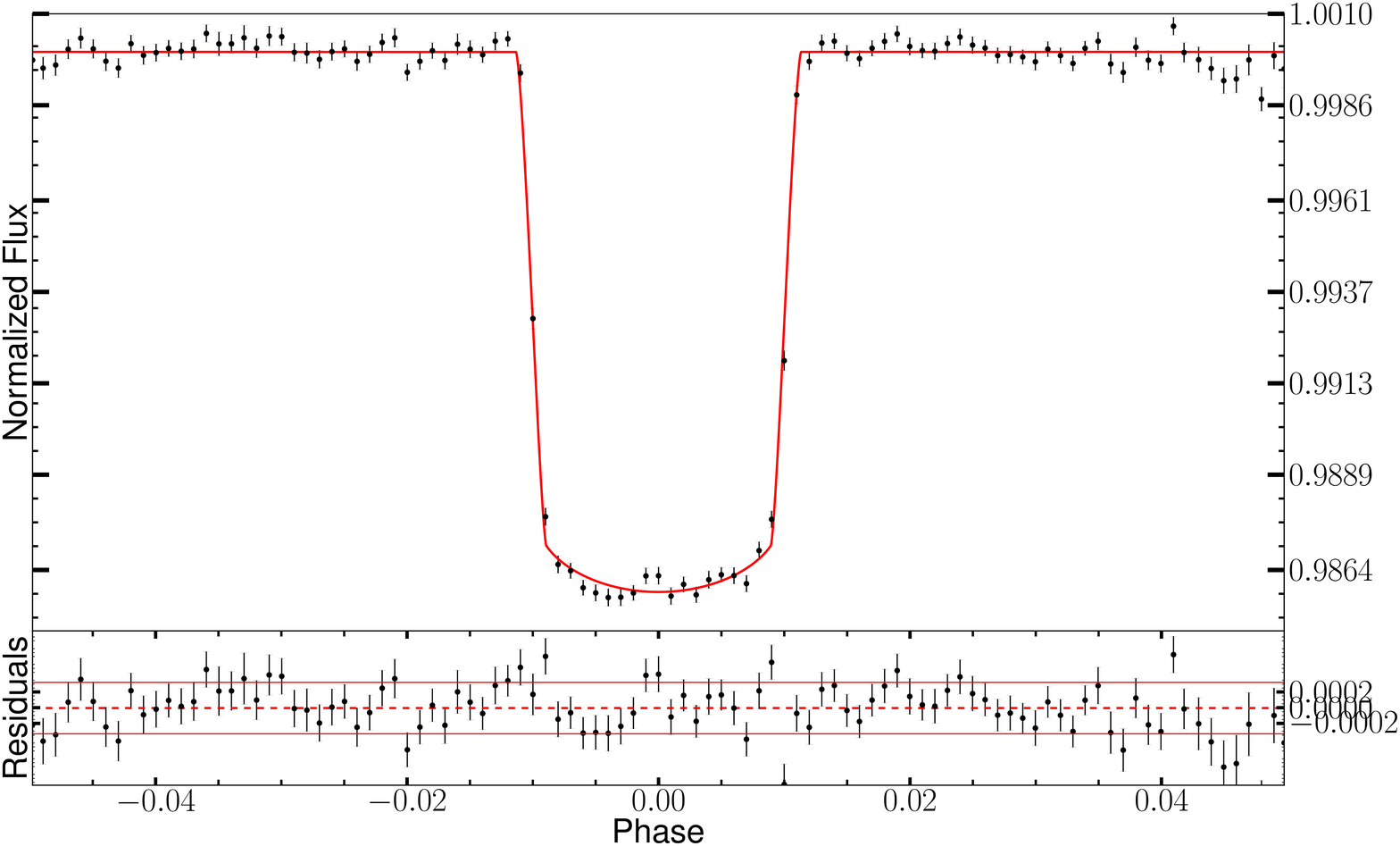}
\vspace{0.7in}
\caption{Phased \& binned transit of GJ1214b at 3.6\,$\mu$m from our two Spitzer transits, plus the transit observed by \citet{desert11}.
\label{fig6}}
\end{figure}

\clearpage

\begin{figure}
\epsscale{.85}
\plotone{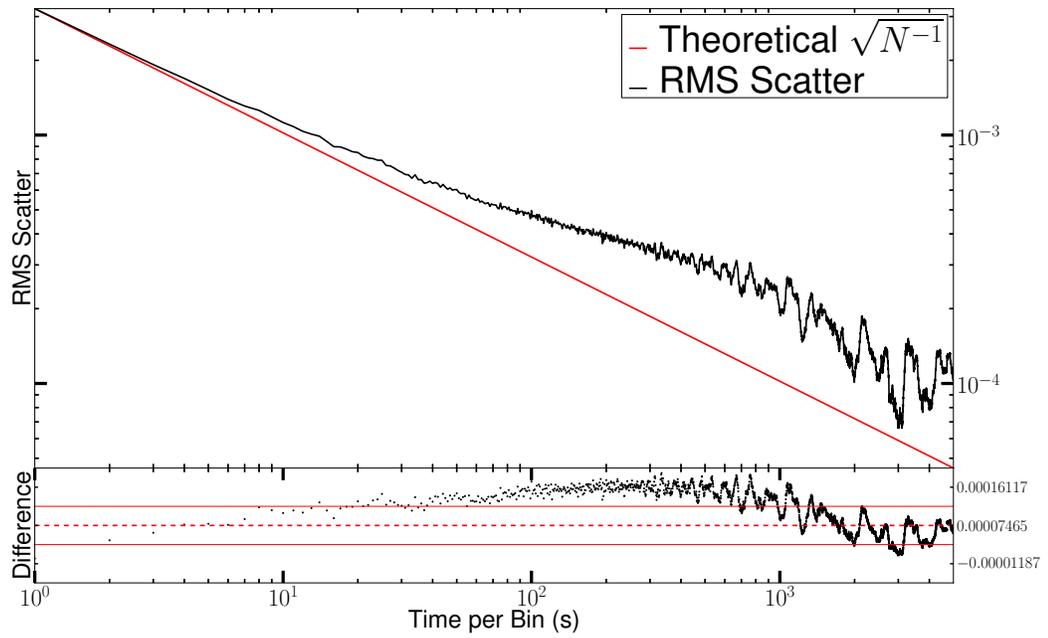}
\vspace{0.7in}
\caption{Standard deviation of the residuals (data minus the `simultaneous'
fit, in units of the stellar flux) for our two transits at 3.6\,$\mu$m
plus the \citet{desert11} transit, versus bin size.  The extra scatter in
this figure, compared to Figure 4 above, is related to the number of data points
binned overall. There are almost an order of magnitude less points at 3.6 $\mu$m than 4.5 $\mu$m.
\label{fig7}}
\end{figure}

\clearpage

\begin{figure}
\epsscale{.85}
\plotone{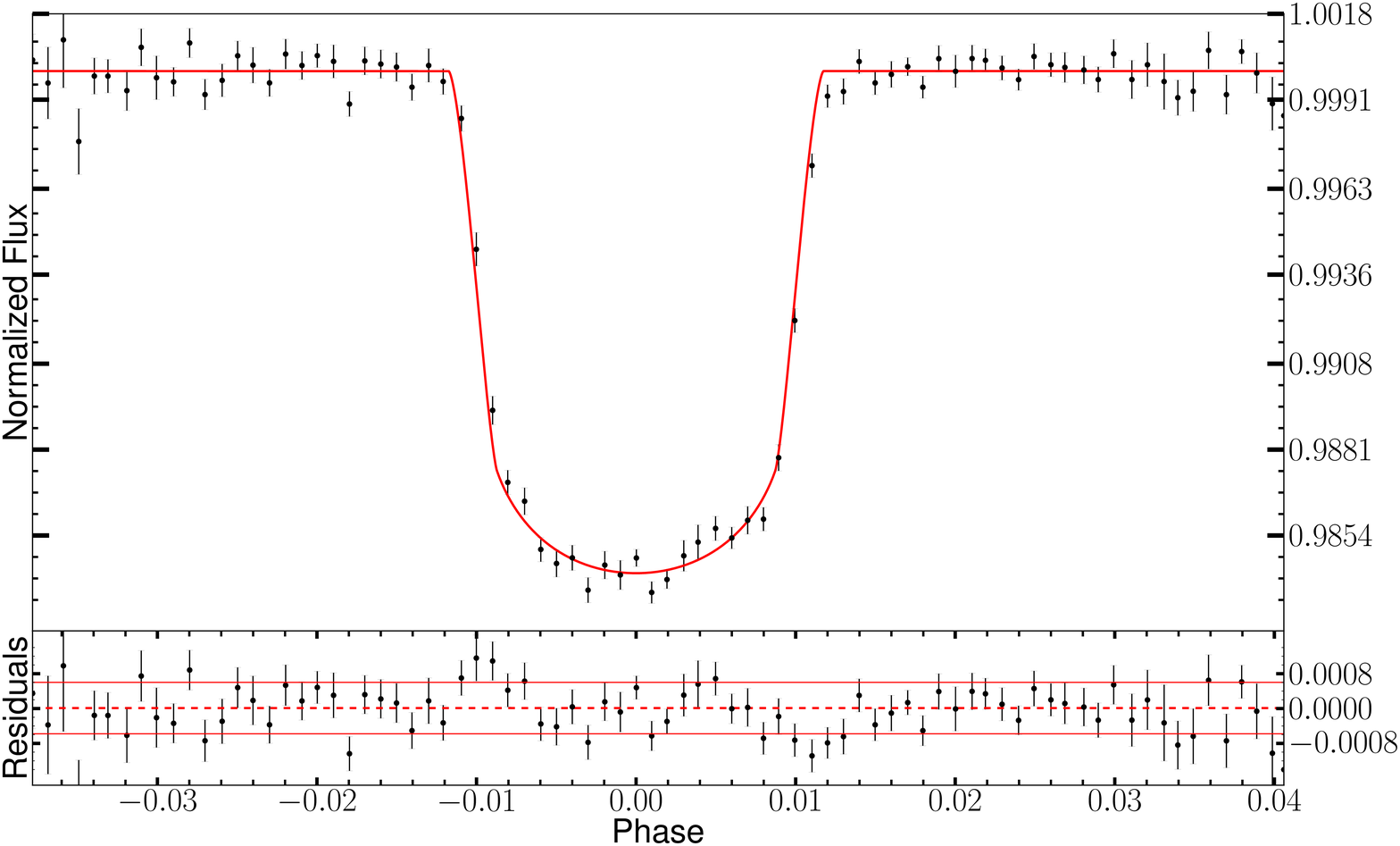}
\vspace{0.7in}
\caption{Phased \& binned transit of GJ1214b in the I+z band, from TRAPPIST, observed by \citet{gillon12}.
\label{fig8}}
\end{figure}

\clearpage

\begin{figure}
\epsscale{.85}
\plotone{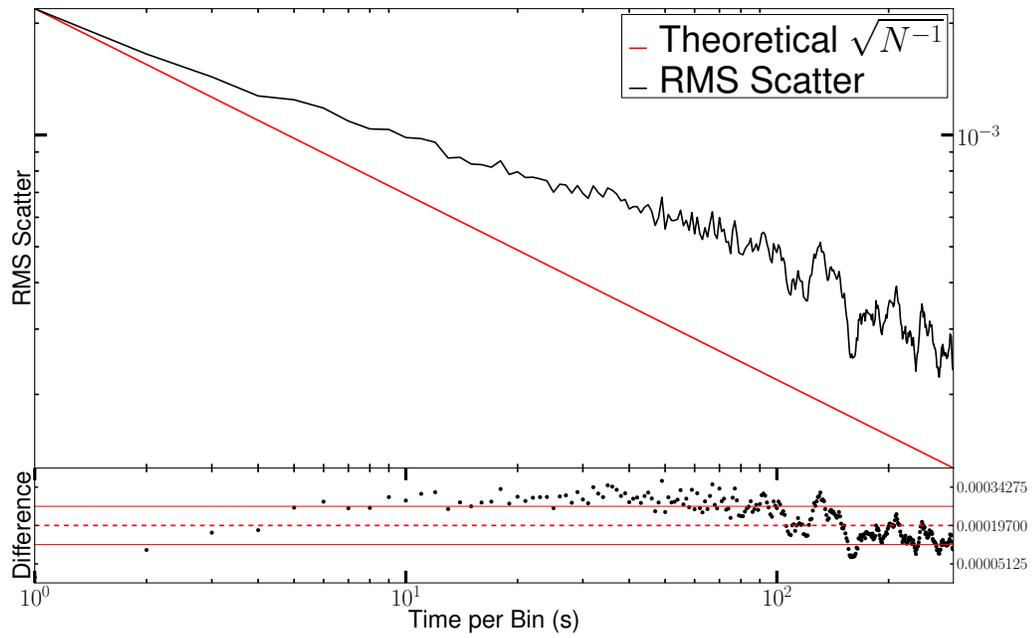}
\vspace{0.7in}
\caption{Standard deviation of the residuals (data minus the `simultaneous'
fit, in units of the stellar flux) for our seven transits in the I+z band on TRAPPIST.  
The extra scatter in this figure, compared to Figure 4 \& 7 above, is related to 
the number of data points binned overall.  We were only able to bin out to 300 
data points ber bin because the most extensive TRAPPIST transit comprised 300 data points.
\label{fig9}}
\end{figure}

\clearpage

\begin{figure}
\epsscale{.85}
\plotone{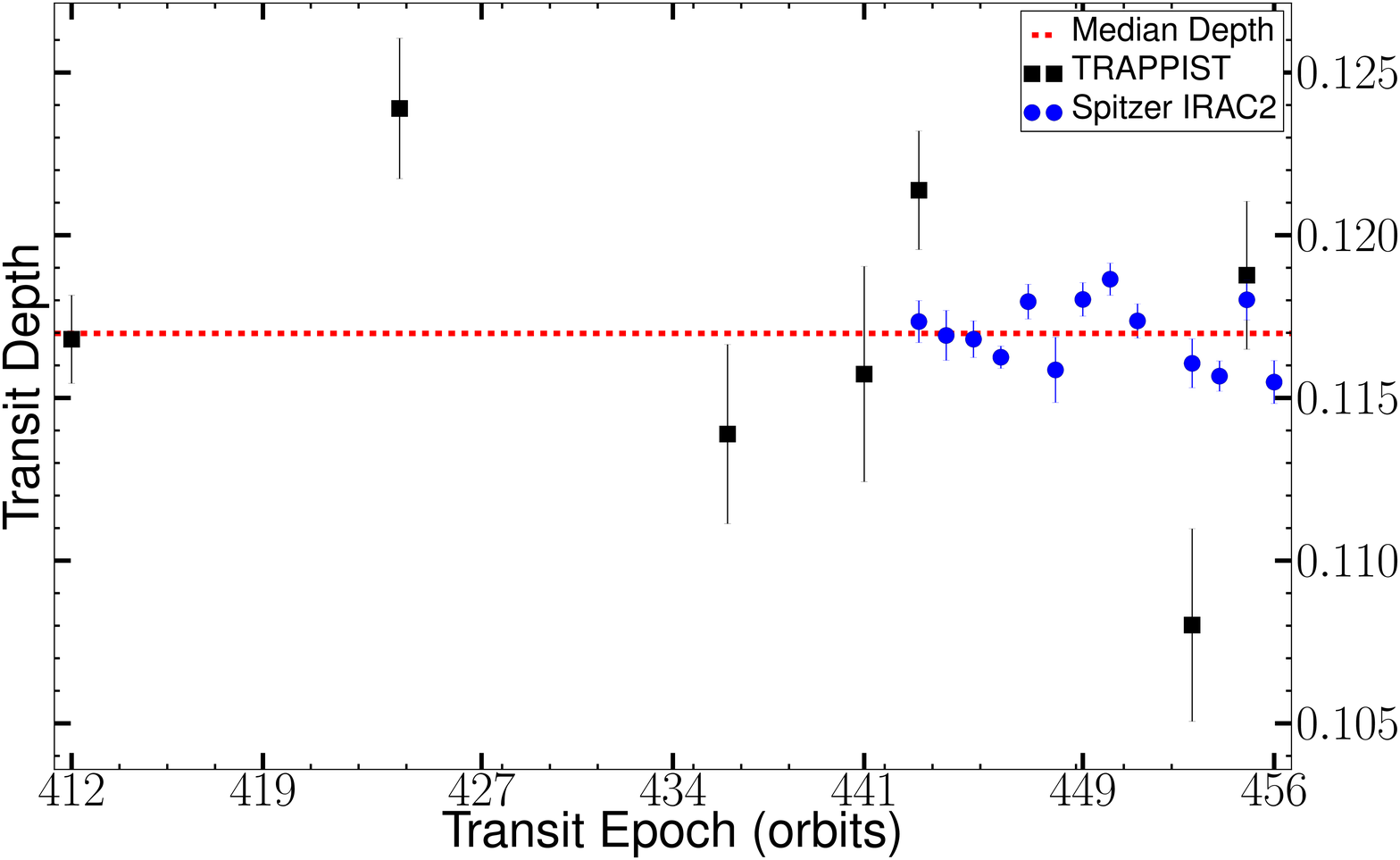}
\vspace{0.7in}
\caption{Ratio of planet to stellar radius versus epoch for our
4.5\,$\mu$m transits, shown in comparison to transits from TRAPPIST.
The red dotted line shows the median value of our 4.5\,$\mu$m transits.
\label{fig10}}
\end{figure}
		   
\clearpage

\begin{figure}
\epsscale{.85}
\plotone{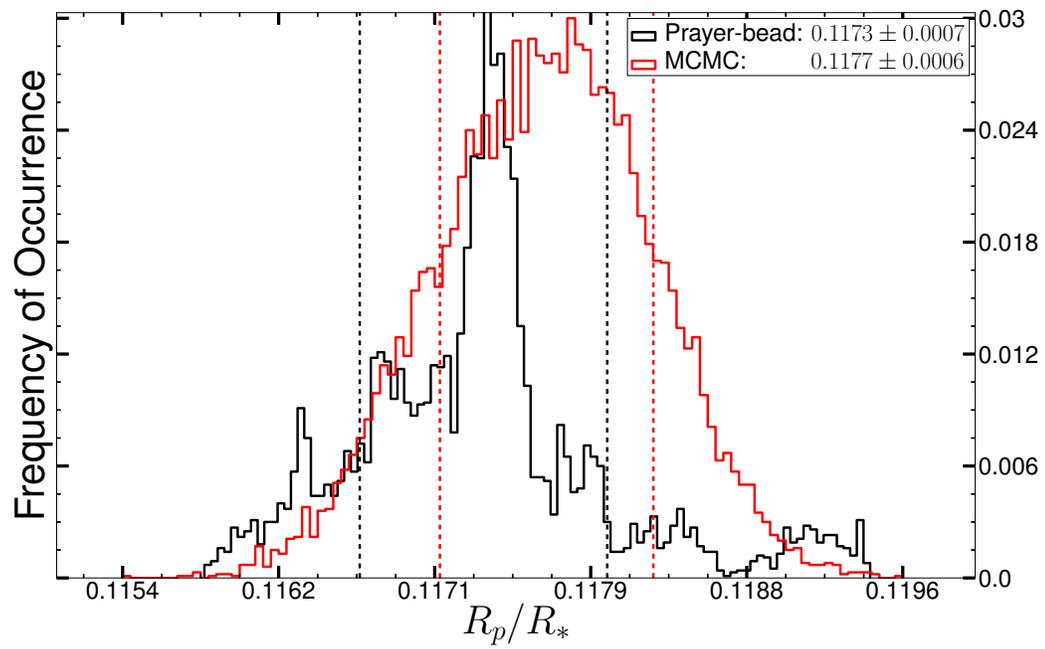}
\vspace{0.7in}
\caption{Histograms (posterior distributions) of ${R_p}/{R_*}$ for the individual fit to the first of our 
4.5\,$\mu$m transits, showing results from both the MCMC and prayer-bead methods;   
with dashed lines for the 1$\sigma$ threshold of each distribution.
\label{fig11}}
\end{figure}

\clearpage

\begin{figure}
\epsscale{.85}
\plotone{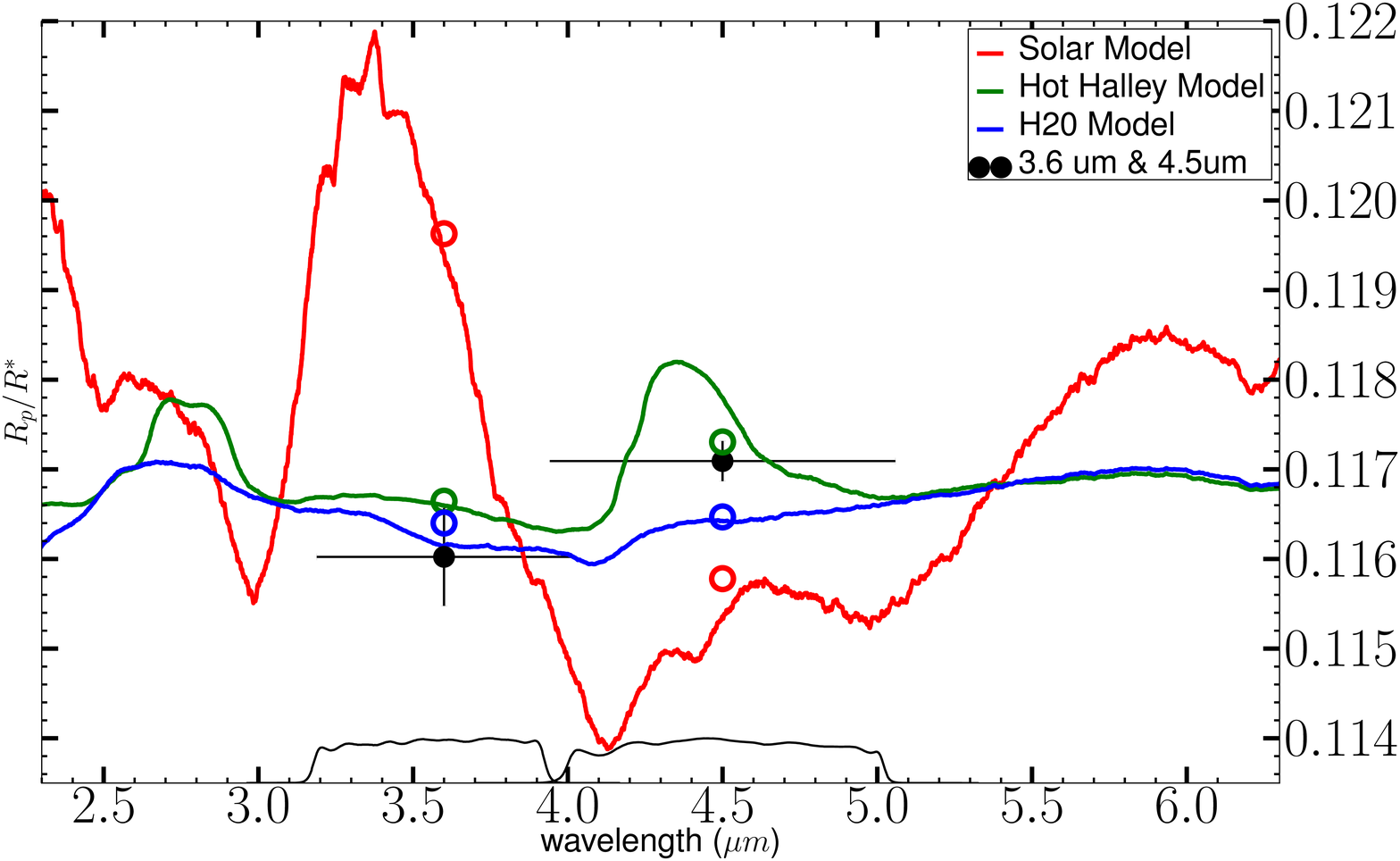}
\vspace{0.7in}
\caption{Comparison between our Spitzer results and three recent models for
the atmosphere of GJ1214b, from \citet{benneke}: a cloudless solar model, 
a comet-like model, and a water vapor model.
\label{fig12}}
\end{figure}

\clearpage

\begin{figure}
\epsscale{.85}
\plotone{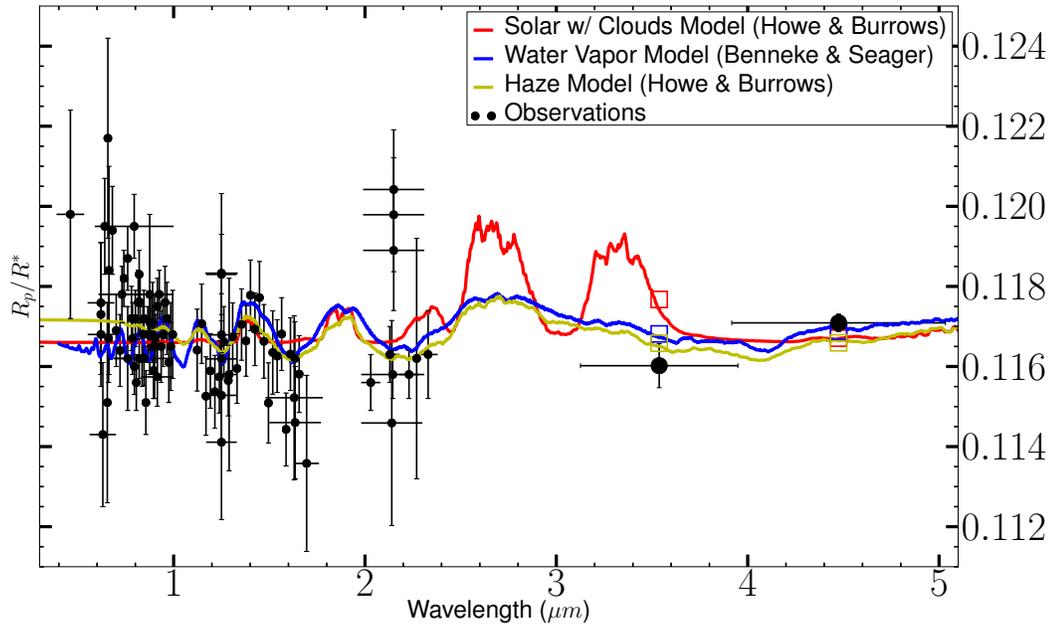}
\vspace{0.7in}
\caption{Comparison between all published observations of GJ1214b,
including our Spitzer and TRAPPIST transits, and recent models. Three
models having the lowest $\chi^2$ are illustrated: a water vapor model
from \citet{benneke} (blue curve), a solar abundance model with very
high cloud tops (at 0.1~mbars, red curve), and a model with 1-percent
water vapor, and a thick tholin haze of 0.1\,$\mu$m particles
extending to very high altitude (1\,$\mu$bar, yellow curve).  The
latter two models are from \citet{howe}.
\label{fig13}}
\end{figure}

\clearpage

\begin{table}\small
\begin{center}
\caption{Results of fitting to 13 individual transits of GJ1214b at 4.5\,$\mu$m, plus the transit observed by \citet{desert11}. \label{tbl-1}}
\begin{tabular}{cccccc}
\tableline
\tableline
 Epoch &  $T_c-2450000 (BJD_{TDB})$  & $i$  & $a/{R_*}$ & ${R_p}/{R_*}$ & $c_0$ \\
\tableline
211 & $5314.214114 \pm 0.000085$ & $88.81 \pm 0.39$ & $14.92 \pm 0.40$ & $0.11726 \pm 0.00105$ & $0.133 \pm 0.074$ \\
443 & $5680.867846 \pm 0.000059$ & $89.30 \pm 0.33$ & $15.51 \pm 0.28$ & $0.11735 \pm 0.00064$ & $0.097 \pm 0.046$ \\
444 & $5682.448272 \pm 0.000067$ & $88.89 \pm 0.37$ & $14.98 \pm 0.39$ & $0.11692 \pm 0.00077$ & $0.208 \pm 0.047$ \\
445 & $5684.028698 \pm 0.000056$ & $89.88 \pm 0.36$ & $15.72 \pm 0.27$ & $0.11681 \pm 0.00057$ & $0.126 \pm 0.041$ \\
446 & $5685.609139 \pm 0.000048$ & $89.11 \pm 0.23$ & $15.24 \pm 0.24$ & $0.11625 \pm 0.00034$ & $0.118 \pm 0.029$ \\
447 & $5687.189632 \pm 0.000065$ & $88.23 \pm 0.26$ & $14.11 \pm 0.41$ & $0.11796 \pm 0.00054$ & $0.059 \pm 0.051$ \\
448 & $5688.769880 \pm 0.000094$ & $89.42 \pm 0.36$ & $15.38 \pm 0.30$ & $0.11586 \pm 0.00100$ & $0.170 \pm 0.067$ \\
449 & $5690.350385 \pm 0.000052$ & $88.34 \pm 0.23$ & $14.33 \pm 0.36$ & $0.11803 \pm 0.00052$ & $0.098 \pm 0.050$ \\
450 & $5691.930794 \pm 0.000067$ & $88.22 \pm 0.25$ & $14.24 \pm 0.43$ & $0.11865 \pm 0.00049$ & $0.033 \pm 0.033$ \\
451 & $5693.511181 \pm 0.000058$ & $88.81 \pm 0.25$ & $14.95 \pm 0.29$ & $0.11737 \pm 0.00053$ & $0.130 \pm 0.046$ \\
453 & $5696.671937 \pm 0.000073$ & $88.68 \pm 0.37$ & $14.75 \pm 0.41$ & $0.11606 \pm 0.00075$ & $0.189 \pm 0.047$ \\
454 & $5698.252258 \pm 0.000047$ & $88.93 \pm 0.28$ & $15.14 \pm 0.30$ & $0.11567 \pm 0.00047$ & $0.056 \pm 0.029$ \\
455 & $5699.832550 \pm 0.000068$ & $88.24 \pm 0.28$ & $14.17 \pm 0.43$ & $0.11802 \pm 0.00062$ & $0.070 \pm 0.058$ \\
456 & $5701.413161 \pm 0.000059$ & $89.61 \pm 0.30$ & $15.41 \pm 0.23$ & $0.11549 \pm 0.00066$ & $0.205 \pm 0.042$ \\
\tableline
\end{tabular}
\end{center}
\end{table}

\begin{table}\small
\begin{center}
\caption{Results of fitting to two individual transits of GJ1214b at 3.6\,$\mu$m, plus the transit observed by \citet{desert11}. \label{tbl-2}}
\begin{tabular}{cccccc}
\tableline
\tableline
 Epoch & $T_c-2450000 (BJD_{TDB})$ & $i$ & $a/{R_*}$ & ${R_p}/{R_*}$ & $c_0$ \\
\tableline
210 & $5312.633724 \pm 0.000072$ & $89.998 \pm 0.324$ & $15.659 \pm 0.160$ & $0.11665 \pm 0.00082$ & $0.200 \pm 0.061$ \\
564 & $5872.096721 \pm 0.000094$ & $89.144 \pm 0.383$ & $15.331 \pm 0.354$ & $0.11619 \pm 0.00121$ & $0.114 \pm 0.065$ \\
565 & $5873.677187 \pm 0.000062$ & $89.221 \pm 0.296$ & $15.418 \pm 0.281$ & $0.11584 \pm 0.00066$ & $0.135 \pm 0.049$ \\
\tableline
\end{tabular}
\end{center}
\end{table}

\begin{sidewaystable}\footnotesize
\begin{center}
\caption{Results of fitting to 7 individual transits of GJ1214b in the I+z-band, observed by TRAPPIST \citep{gillon12}.  \label{tbl-3}}
\begin{tabular}{ccccccc}
\tableline
\tableline
  Epoch & $T_c -2450000 (BJD_{TDB})$  & $i$  & $a/{R_*}$ & ${R_p}/{R_*}$ & $c_0$ & $c_1$\\
\tableline
412 & $5631.875656 \pm 0.000092$ & $88.035 \pm 0.365$ & $13.743 \pm 0.443$ & $0.11680 \pm 0.00135$ & $0.237 \pm 0.094$ & $0.073 \pm 0.177$ \\
424 & $5650.840034 \pm 0.000152$ & $87.805 \pm 0.384$ & $13.215 \pm 0.515$ & $0.12389 \pm 0.00216$ & $0.331 \pm 0.176$ & $0.110 \pm 0.220$ \\
436 & $5669.805222 \pm 0.000110$ & $88.990 \pm 0.436$ & $14.945 \pm 0.562$ & $0.11389 \pm 0.00275$ & $0.422 \pm 0.170$ & $0.267 \pm 0.345$ \\
441 & $5677.706878 \pm 0.000170$ & $88.965 \pm 0.412$ & $14.923 \pm 0.550$ & $0.11573 \pm 0.00331$ & $0.428 \pm 0.198$ & $0.008 \pm 0.370$ \\
443 & $5680.867699 \pm 0.000131$ & $88.336 \pm 0.397$ & $14.049 \pm 0.486$ & $0.12138 \pm 0.00182$ & $0.012 \pm 0.153$ & $0.602 \pm 0.298$ \\
453 & $5696.672075 \pm 0.000135$ & $89.873 \pm 0.671$ & $14.932 \pm 0.741$ & $0.10802 \pm 0.00296$ & $0.646 \pm 0.142$ & $0.141 \pm 0.242$ \\
455 & $5699.832535 \pm 0.000143$ & $88.360 \pm 0.460$ & $14.516 \pm 0.641$ & $0.11877 \pm 0.00228$ & $0.343 \pm 0.138$ & $0.005 \pm 0.268$ \\
\tableline
\end{tabular}
\end{center}
\end{sidewaystable}

\begin{sidewaystable}\footnotesize
\begin{center}
\caption{Results of fitting to composite transits, i.e.: simultaneous fits, averaging individual transits, and fitting to phased \& binned
transits.  
\label{tbl-4}}
\begin{tabular}{ccccccc}
\tableline
\tableline
Method / Channel & $T_c-2450000 (BJD_{TDB})$ & $i$  & $a/{R_*}$ & ${R_p}/{R_*}$ & $c_0$ & $c_1$ \\
\tableline
Simultaneous 3.6\,$\mu$m 			& $4966.524895 \pm 0.000047$ & $89.566 \pm 0.139$ & $15.607 \pm 0.104$ & $0.11607 \pm 0.00030$ & $0.156 \pm 0.022$ & fixed at 0.0 		\\
Simultaneous 4.5\,$\mu$m	 		& $4966.524918 \pm 0.000030$ & $88.794 \pm 0.061$ & $15.049 \pm 0.095$ & $0.11710 \pm 0.00017$ & $0.110 \pm 0.014$ & fixed at 0.0	 	\\
Simultaneous TRAPPIST I+z     		& $4966.524881 \pm 0.000063$ & $88.240 \pm 0.230$ & $14.011 \pm 0.288$ & $0.11873 \pm 0.00111$ & $0.320 \pm 0.097$ & $0.187 \pm 0.176$  \\
Averaged individual 3.6\,$\mu$m 	& $4966.524898 \pm 0.000076$ & $89.469 \pm 0.221$ & $15.564 \pm 0.080$ & $0.11616 \pm 0.00019$ & $0.148 \pm 0.021$ & fixed at 0.0 		\\
Averaged individual 4.5\,$\mu$m		& $4966.524925 \pm 0.000027$ & $88.803 \pm 0.127$ & $15.090 \pm 0.137$ & $0.11699 \pm 0.00026$ & $0.110 \pm 0.015$ & fixed at 0.0 		\\
Averaged individual TRAPPIST I+z 	& $4966.524968 \pm 0.000077$ & $88.437 \pm 0.057$ & $14.207 \pm 0.239$ & $0.11793 \pm 0.00182$ & $0.324 \pm 0.068$ & $0.146 \pm 0.073$	\\
Phased \& binned 3.6\,$\mu$m 		& $4966.524156 \pm 0.000043$ & $89.469 \pm 0.143$ & $15.547 \pm 0.091$ & $0.11602 \pm 0.00055$ & $0.158 \pm 0.035$ & fixed at 0.0 		\\
Phased \& binned 4.5\,$\mu$m 		& $4966.524297 \pm 0.000016$ & $88.439 \pm 0.061$ & $14.442 \pm 0.093$ & $0.11709 \pm 0.00022$ & $0.128 \pm 0.018$ & fixed at 0.0 		\\
Phased \& binned TRAPPIST I+z 		& $4966.524122 \pm 0.000079$ & $88.314 \pm 0.177$ & $14.053 \pm 0.259$ & $0.11803 \pm 0.00079$ & $0.263 \pm 0.107$ & $0.316 \pm 0.149$	\\
\tableline
\end{tabular}
\end{center}
\end{sidewaystable}

\begin{table}\footnotesize
\begin{center}
\caption{Summary of results for $R_{p}/R_{*}$ derived at different
wavelengths using different methods. The phased \& binned transit
results for $R_p/R_{*}$ are given for several cases: our best-fit
orbital parameters ($a/R_{*}$,~$i$), as well as $R_p/R_{*}$ when we
force the orbital parameters to have the values determined by
\citet{bean11} (14.97,~88.94) and \citet{berta12} (15.31,~89.30),
respectively. 
\label{tbl-5}}
\begin{tabular}{lc}
Method & ${R_p}/{R_*}$ \\
\tableline\\
Spitzer 4.5\,$\mu$m\\
\tableline\tableline
Simultaneous 			& $0.1171 \pm 0.0002$ \\
Averaged Individual		& $0.1170 \pm 0.0003$ \\
\tableline \\
Phased \& binned
Best-fit orbital		& $0.1171 \pm 0.0002$ \\
Bean orbital            & $0.1161 \pm 0.0003$ \\
Berta orbital           & $0.1160 \pm 0.0003$ \\
\tableline\\
Spitzer 3.6\,$\mu$m\\
\tableline\tableline
Simultaneous 			& $0.1161 \pm 0.0003$ \\
Averaged Individual     & $0.1162 \pm 0.0001$ \\
\tableline\\
Phased \& binned\\
Best-fit orbital		& $0.1160 \pm 0.0006$ \\
Bean orbital            & $0.1163 \pm 0.0002$ \\
Berta orbital           & $0.1161 \pm 0.0002$ \\
\tableline\\
TRAPPIST I+z band\\
\tableline\tableline
Simultaneous			& $0.1187 \pm 0.0011$ \\
Averaged Individual     & $0.1179 \pm 0.0018$ \\
\tableline\\
Phased \& binned\\
Best-fit orbital		& $0.1180 \pm 0.0005$ \\
Bean orbital            & $0.1177 \pm 0.0005$ \\
Berta orbital           & $0.1172 \pm 0.0005$ \\
\tableline
\end{tabular}
\end{center}
\end{table}



\clearpage




\end{document}